\documentclass[a4paper]{article}
\usepackage{graphicx}

\begin{document}
\title{The interpretation of quantum mechanics: where do we stand?}

\author{ GianCarlo Ghirardi\footnote{e-mail: ghirardi@ts.infn.it}\\
{\small Department of Theoretical Physics of the University of Trieste, and}\\
{\small the Abdus Salam International Centre for Theoretical Physics,
Trieste, Italy, and}\\{\small the Istituto Nazionale di Fisica Nucleare.}}

\date{}
\maketitle

\begin{abstract}
We reconsider some important foundational problems of quantum mechanics. After reviewing the measurement problem and discussing its unavoidability, we analyze some proposals to overcome it. This analysis  leads us to reconsider the current debate on our best theory, i.e. quantum mechanics itself. We stress that, after the remarkable interest and the many efforts which have lead, in the last years of the past century, to a revival of  the subject, and, more important, to new interesting results, we are now witnessing a re-emergence of the vague and unprofessional positions which have characterized the debate in the second quarter of the XXth century. In particular
 we consider as extremely serious the fact that a completely mistaken position concerning the real meaning of Bell's theorem seems to have been taken by many scientists in the field.
 \end{abstract}

\section{Introduction}
As well known, almost one century after its formulation and in spite of its unprecedented predictive successes in accounting for physical processes, quantum mechanics is still at the centre of a lively debate. Everybody knows very well the disputes between some of the great founding fathers of the theory, from N Bohr to A Einstein to E Schr\"{o}dinger to W Heisenberg, which have marked the twenties and the thirties of the past century. The debate has been less heated in the subsequent years, perhaps because, as stated by M Gell-Mann in his Nobel acceptance speech in 1976:
\begin{quote}
{\it Niels Bohr brainwashed a whole generation of physicists into believing that the problem (of the interpretation of the theory) had been solved fifty years ago.}
\end{quote} 

Actually I remember well that, at the beginning of the sixties when I started my scientific carreer, to work on foundational issues was considered by a great part of the scientific community  a loss of time,  a choice to pay more attention to (irrelevant) philosophical issues than to precise scientific problems. Luckily enough, the sixties were also the years in which another deep thinker, John S Bell,  by deriving the celebrated inequality that bears his name \cite{Bell1}, has given a tremendous imput to our understanding of reality by making clear that nonlocal features  characterize most natural processes. However, this great scientist did not limit his considerations to the peculiar aspects of the theory stemming from one of its most surprising features, entanglement, but has devoted many important papers to give voice to his unsatisfaction with the conceptual and logical status of the theory, in particular with  the so-called measurement or macro-objectification problem.
His immense prestige has pushed many physicists to reconsider such a problem and during the fall of the past century there has been a re-flourishing of foundational investigations.

In this paper we will first of all concentrate our attention on the macro-objectification problem and on some recent  proposals to overcome it, and then we will briefly review and analyze the  positions which characterize the present debate on this problem.

\section{The macro-objectification problem}

The problem we are interested in is usually presented by making reference to the so-called von Neumann ideal measurement scheme. Synthetically it goes as follows. 

Suppose one is interested in {\it measuring} a physical observable  $F$ of a microscopic system $S$. Let us denote as $\mathcal F$ the (self-adjoint) operator associated to it, as $ f_{i}$ its eigenvalues and as $\vert\phi_{i}\rangle$ the corresponding eigenvectors. Obviously, since $S$ is microscopic we have not direct access to it and therefore, to actually perform the measurement, we assume, with von Neumann, that we have at our disposal a macroscopic quantum system in a ``ready state" $\vert A_{R}\rangle$ which we put into interaction with $S$ and that the system apparatus interaction which lasts for a certain time interval is such that the initial state $\vert\phi_{i}\rangle\otimes\vert A_{R}\rangle$ evolves into the final state $\vert\phi_{i}\rangle\otimes\vert A_{i}\rangle$. Here the final states of the apparatus $\vert A_{i}\rangle$ are orthogonal and macroscopically different. In this way, by looking at the final state of the apparatus   we can deduce in which of the eigenstates of $\mathcal F$  the microsystem was before the measurement.

Now comes the problem: quantum mechanics is a linear theory, implying that if various states are possible states for a system $S$ then also any normalized linear combination of them   is a possible state for it. Moreover, the evolution law is linear, so that a linear superposition of  states evolves into the same linear superposition of the evolved of such states. In formulae and with reference to our example:

\begin{equation}
\vert\phi_{i}\rangle\otimes\vert A_{R}\rangle\rightarrow\vert\phi_{i}\rangle\otimes\vert A_{i}\rangle \Rightarrow \sum_{i} c_{i}\vert\phi_{i}\rangle\otimes\vert A_{R}\rangle\rightarrow\sum_{i}c_{i}\vert\phi_{i}\rangle\otimes\vert A_{i}\rangle,
\end{equation} 

\noindent where the simple arrow denotes the time evolution and the double one denotes the implication.

As Eq. (1) clearly shows, in the final state  a superposition of macroscopically different states appears. Now, what meaning whatsoever can we attribute to a state which is a superposition (and not a statistical mixture) of such states? Actually the states $\vert A_{i}\rangle$ of the apparatus might correspond to different locations of a macroscopic pointer. How can then a macroscopic object be in a state which does not correspond to a  macroscopically definite position?

We all know the answer that the so-called ortodox interpretation of quantum mechanics gives to our question: wave packet reduction takes place during the process and the final state is not the one appearing at the end of Eq.(1), but, with a probability given by $\vert c_{j}\vert^{2}$,  the state $\vert\phi_{j}\rangle\otimes\vert A_{j}\rangle$ of the superposition.

This assumption is fundamentally inconsistent since it contemplates two contradictory evolution laws for physical systems: one, which is linear and deterministic, governing the evolution of microsystems, and one, which is nonlinear and stochastic, holding when macroscopic systems enter into play. Why should macrosystems not to be governed by quantum mechanics? Are they not built up of elementary constituents (electrons, protons, neutrons) which are subjected to the laws of quantum mechanics? But this is not the whole story. In fact, on the one hand, we know very well that there are macroscopic systems whose properties can be understood only in terms of quantum theory, while, on the other, even if we would be inclined to accept two different evolution principles, we cannot avoid to recognize that there is nothing, absolutely nothing, in the theory which allows us to locate the split identifying which systems obey  the linear equation and which ones do not. As lucidly summarized by Bell \cite{Bell2}:

\begin {quote}
{\it There is a fundamental ambiguity in quantum mechanics, in that nobody knows exactly what it says about any particular situation, for nobody knows exactly where the boundary between the wavy quantum world and the world of particular events is located.}
\end{quote}

To conclude this section we remark that the above ideal scheme of measurement (such measurement processes are usually denoted as `of the first kind')  is an extremely idealized one: in practice, the final apparatus states are not strictly orthogonal, we cannot control all degrees of freedom characterizing them, a macroscopic object cannot be isolated from the environment, no apparatus is exempt from malfunctioning, etc. Some scientists maintain that it is just resorting to such an idealized description  as the one of Eq.(1) which leads to the inconsistencies we have just discussed. As a paradigmatic example we will mention a sentence \cite{Primas} by H. Primas:

\begin{quote}
{\it If you really want to discuss measurements of the first kind you may try your luck. But remember that measurements of the first kind are unrealistic and completely irrelevant for experimental science. Replace in all books and philosophical treatises on quantum mechanics the word `measurement' by the proper expression `measurement of the first kind' and add a footnote: measurements of the first kind are idealizations which never play any role in experimental sciences.}
\end{quote}

Our next step is to prove that this position is plainly wrong and that the macro-objectification problem cannot be avoided. To this purpose we will briefly recall the argument of a recent paper \cite{Bassi} in which the problem has been tackled in its full generality.

\section{A completely general measurement scheme}

To exhibit an explicit proof of the last statement of the previous section, let us consider a microsystem which we assume can be prepared in two orthogonal states  $\vert u\rangle$ and $\vert d\rangle$ (typically we can think of the two spin states of a spin $1/2$ particle corresponding to a definite value of $\sigma_{z}$), and in  their equal amplitudes superposition $\vert u+d\rangle=\frac{1}{\sqrt{2}}[\vert u\rangle+\vert d\rangle]$. We consider also a statistical ensemble $\mathcal E \{\vert A_{R},\alpha\rangle,p(\alpha)\}$ of apparata, devised to measure precisely the observable $\sigma_{z}$. We have resorted to  an ensemble of apparata just because we cannot have full control of all their degrees of freedom. The state $\vert A_{R},\alpha\rangle$  corresponds to the `ready state' of the apparatus, let us say one for which the macroscopic pointer of the apparatus `points,  at 0', while the index $\alpha$ distinguishes situations differing for uncontrollable degrees of freedom such as the precise state of the molecules of the pointer, and so on\footnote {Note that we can include in $\alpha$ also degrees of freedom which are external to the apparatus, typically, those referring to the environment. }. The quantities $p(\alpha)$ are the weights of the various subensembles of the statistical ensemble characterized by the macroproperty $A_{R}$. Completely in general, when we will resort to expressions of the type $\vert X, \mu\rangle$ we intend to indicate a situation of the macro-apparatus corresponding to what we perceive\footnote{It goes without saying that all statements concerning the position of the pointer do not make reference to a situation in which the position of the pointer is a geometrical point, but to one in which it lays in a quite narrow interval on a graduated scale.} as `the pointer point at X' and all other relevant variables are summarized by the symbol $\mu$. Let us denote as $\mathcal X$ the set of all such states for all possible values of $\mu$. We will not require orthogonality of states corresponding to macroscopically different situations, but, obviously, we have to impose that two such states are, to a large extent, ``distinguishable", a fact that we will characterize by imposing:
\begin{equation}
inf_{\mu \in \mathcal X, \nu\in \mathcal Y}\parallel\vert X,\mu\rangle-\vert Y,\nu\rangle\parallel\geq\sqrt{2}-\eta,\;\;\; \eta\ll1.
\end{equation}

Let us go on with our assumptions. We put the system in one of the two states $\vert u\rangle$ or $\vert d\rangle$ into interaction with the specific apparatus in the state $\vert A_{R},\alpha\rangle$, and we assume that the quantum evolution during the measurement is governed by a unitary operator $U(t_{i},t_{f})$. Let us denote as $\vert F,u,\alpha\rangle$ and $\vert F,d,\alpha\rangle$ the final states obtained when we trigger the apparatus in the state $\vert A_{R},\alpha\rangle$ by the microstates $\vert u\rangle$ and $\vert d\rangle$, respectively:
\begin{equation}
U(t_{i},t_{f})[\vert u\rangle\otimes\vert A_{R},\alpha\rangle]=\vert F,u,\alpha\rangle;\;\;\;U(t_{i},t_{f})[\vert d\rangle\otimes\vert A_{R},\alpha\rangle]=\vert F,d,\alpha\rangle.
\end{equation}  

Let us now consider some physically meaningful sets:

\begin{equation}
J_{\mathcal U}^{-}=\{\alpha: \vert F,u, \alpha\rangle \notin \mathcal U\};\;\;\;J_{\mathcal D}^{-}=\{\alpha: \vert F,d, \alpha\rangle \notin \mathcal D\},
\end{equation} 
and their complements $J_{\mathcal U}^{+}=\mathcal C J_{\mathcal U}^{-} $, $J_{\mathcal D}^{+}=\mathcal C J_{\mathcal D}^{-} $.   
The physical meaning of these sets should be clear. For instance $J_{\mathcal U}^{-}$ represents the subset of those apparata (actually the set of the $\alpha$'s) for which, in spite of having being triggered by the microstate $\vert u\rangle$, their final statevector does not belong to   the set $\mathcal U$, i.e. the one for which the pointer `points at U'. For $\alpha$ in such a set the apparata did not register correctly the fact that  the system we are interested in is in the state $\vert u\rangle$. The meaning of the other three sets, is obvious.
We can now formulate our final assumption. By taking advantage of the fact that the function $p(\alpha)$ is a probability measure on the set on which the index $\alpha$ runs, we can define a measure by the formula:
\begin{equation}
\mu \{\mathcal X\}=\int_{\alpha\in \mathcal X} p(\alpha) d\alpha.
\end{equation}
We now make a natural reliability assumption concerning our ensemble of apparatuses:
\begin{equation}
\mu \{J_{\mathcal U}^{-}\}<\epsilon, \;\;\;\mu \{J_{\mathcal D}^{-}\}<\epsilon\;\;\;\; i.e.\;\;\;\;\mu \{J_{\mathcal U}^{+}\}>1-\epsilon, \;\;\;\mu \{J_{\mathcal D}^{+}\}>1-\epsilon.  
\end{equation}
Note that these relations imply that the set of all $\alpha$'s for which the apparatuses give the right answer both when triggered by $\vert u\rangle$ or by $\vert d\rangle$ has a measure:

\begin{equation}
\mu\{J_{\mathcal U}^{+}\cap J_{\mathcal D}^{+}\}>1-2\epsilon,\;\;\;\epsilon\ll 1.
\end{equation}
which is near to 1.

Having made precise our assumptions and notation we can now proceed as follows. Let us trigger those apparatuses for which $\alpha$ belongs to $J_{\mathcal U}^{+}\cap J_{\mathcal D}^{+}$ (i.e. those which register correctly both the state $\vert u\rangle$ as well as the state $\vert d\rangle$) by the state $\vert u+d\rangle$. The linear nature of quantum mechanics implies, from Eq. (3):

\begin{equation}
U(t_{i},t_{f})\frac{1}{\sqrt{2}}[\vert u+d\rangle\otimes\vert A_{R},\alpha\rangle\equiv\frac{1}{\sqrt{2}} ]U(t_{i},t_{f})[\vert u\rangle+\vert d\rangle]\otimes\vert A_{R},\alpha\rangle=\frac{1}{\sqrt{2}}[\vert F,u,\alpha\rangle+\vert F,d,\alpha\rangle].
\end{equation}

Let us now raise the question: does the final state of Eq.(8) belong to the set $\mathcal U$ of states? The answer is no, because its distance from the state $\vert F,d,\alpha\rangle$ is:
\begin{equation}
\parallel \vert F, u+d,\alpha\rangle-\vert F,d,\alpha\rangle\parallel=\parallel\frac{1}{\sqrt{2}}\vert F,u,\alpha\rangle+(1-\frac{1}{\sqrt{2}})\vert F,d,\alpha\rangle\parallel\le \frac{1}{\sqrt{2}}+1-\frac{1}{\sqrt{2}}=1, 
\end{equation}
while, according to Eq.(2) a state belonging to $\mathcal U$ must have a distance almost equal to $\sqrt{2}$ from any state belonging to the set $\mathcal D$ of states corresponding to a macroscopically and perceptively different situation from the one which is, by assumption, associated to the perception `the pointer points at U'. The same argument can be followed to prove that the final state cannot belong to $\mathcal D$ or to any other set of states corresponding to a definite position of the pointer. 

The conclusion should be obvious. Having assumed that we have an ensemble $\mathcal E \{\vert A_{R},\alpha\rangle, p(\alpha)\}$ of macro-apparata which allows us  reliably (but not perfectly) to know whether the microstate triggering them is either $\vert u\rangle$ or $\vert d\rangle$, then the large majority of such apparata (i.e. all those characterized by an $\alpha$ belonging to the set $ J_{\mathcal U}^{+}\cap J_{\mathcal D}^{+}$, whose measure is near to 1), when triggered by the superposition $\vert u+d\rangle$ ends up in a state which cannot correspond to any definite position-perception concerning the macroscopic pointer. The macro-objectification problem cannot be overcome invoking measurement processes more realistic than the von Neumann one: it is the very possibility of measuring an observable with an acceptable level of reliability and the assumption of the unrestricted validity of quantum mechanics which imply that the same micro-macro interaction leads almost always to a perceptually nondefinite situation when the triggering state is the superposition of different eigenstates of the observed quantity.

\section{Decoherence}

A quite popular proposal which has been put forward to overcome the macro-objectification problem is based on the consideration of the unavoidable and uncontrollable interaction of any macroscopic body with its environment. A  sketchy way of summarizing this proposal is the following. Let us consider a state which is a linear superposition of two macroscopically different states of a macro-object $S$:
\begin{equation}
\vert\Psi^{(sys)}\rangle=\alpha\vert M\rangle+\beta \vert N\rangle.
\end{equation}

One then takes into account the above mentioned unavoidable interactions of $S$ with the environment, which we suppose to be   described by the state $\vert E_{0}\rangle$ at the initial time. Such interactions and the subsequent evolution lead  to a state of the system+environment which we will express as:

\begin{equation}
\vert\Psi^{(sys+env)}\rangle=\alpha\vert M\rangle\otimes\vert E_{M}\rangle+\beta \vert N\rangle\otimes\vert E_{N}\rangle.
\end{equation}

Now we have to take into account two important facts. The states $\vert E_{M}\rangle$ and $\vert E_{N}\rangle$ appearing in the above equation turn out to be practically orthogonal and they are essentially out of the control of the experimenter. The second fact is obvious, the first is made understandable by taking into account that, e.g., the interaction of a macrosystem which is, let us say, located in a certain position (when the state is $\vert M\rangle$) interacting with a single molecule of the environment can easily throw it into an excited state which is orthogonal to the state of the same molecule when it has not interacted with the macrosystem since it is differently located (as, we suppose, is the case for the state $\vert N\rangle$). But this  is not the whole story; even if for the two macroscopically different states of the macrosystem no one of the microsystems of the environment is in   one of a pair of orthogonal states, for sure many microsystems will be in different states in the two cases. Such states, being different, have a scalar product which is smaller than 1. If one then takes into account that we can easily have, let us say, an Avogadro number of such states (corresponding to the many particles which have interacted with the macrosystem), one easily realizes that the two states $\vert E_{M}\rangle$ and $\vert E_{N}\rangle$ are, {\it de facto}, orthogonal.

As already stated, however, the experimenter has no control over the states of the particles of the environment: he is only  interested in the physics of the macrosystem he is studying. Now, quantum mechanics teaches us that to describe the properties of a subsystem of a composite system, the most efficient and formally direct way is the one of resorting to the statistical operator formalism and to take the partial trace on the degrees of freedom we are disregarding. Let me summarize the steps which are involved using both the language of the statevectors and of the statistical operators:

\begin{eqnarray}
\vert \Psi^{(sys)}\rangle & \rightarrow &  \vert \Psi^{(sys+env)}\rangle\\ \nonumber
\rho^{(sys)}=\vert \Psi^{(sys)}\rangle\langle\Psi^{(sys)}\vert & \rightarrow & \rho^{(sys+env)}=\vert \Psi^{(sys+env)}\rangle\langle\Psi^{(sys+env)}\vert 
\end{eqnarray}

We recall now the explicit form of $\rho^{(sys+env)}$ following from Eq.(11):
\begin{eqnarray}
\rho^{(sys+env)}=\vert \alpha\vert^{2} \vert M\rangle\otimes\vert E_{M}\rangle\langle E_{M}\vert\otimes\langle M\vert & + & \vert \beta\vert^{2} \vert N\rangle\otimes\vert E_{N}\rangle\langle E_{N}\vert\otimes\langle N\vert+\\ \nonumber + \alpha\beta^{*} \vert M\rangle\otimes\vert E_{M}\rangle\langle E_{N}\vert\otimes\langle N\vert & + & \alpha^{*}\beta \vert N\rangle\otimes\vert E_{N}\rangle\langle E_{M}\vert\otimes\langle M\vert.
\end{eqnarray}
If we now evaluate the reduced statistical operator, taking into account the orthogonality of the states of the environment we have:

\begin{equation}
\rho^{(Red,sys)}\equiv Tr^{(env)}\rho^{(sys+env)}=\vert \alpha\vert^{2} \vert M\rangle\otimes\vert E_{M}\rangle\langle E_{M}\vert\otimes\langle M\vert+\vert \beta\vert^{2} \vert N\rangle\otimes\vert E_{N}\rangle\langle E_{N}\vert\otimes\langle N\vert.
\end{equation}

Such a statistical operator is the one associated to the statistical mixture of the states $\vert M\rangle$ and $\vert N\rangle$ with statistical weights $\vert\alpha\vert^{2}$ and $\vert\beta\vert^{2}$, a fact which would render legitimate the replacement of the original pure state of the system with the corresponding mixture:
\begin{equation}
\vert\Psi^{(sys)}\rangle=\alpha\vert M\rangle+\beta \vert N\rangle\Rightarrow\mathcal E\{\vert M\rangle,\vert N\rangle;  \vert\alpha\vert^{2},\vert\beta\vert^{2}\},
\end{equation}
with obvious meaning of the symbols.

The argument seems appealing and actually puts into evidence one relevant fact, i.e., that to experimentally distinguish the pure state from the statistical mixture is an extremely difficult task. But if one looks at it from the conceptual point of view one immediately understands that it is seriously misleading for at least two reasons:
\begin{itemize}
\item In the previous section we have proved that actually linear superpositions of macroscopically different states occur and that their occurrence leads to a deadlock. To replace them with a statistical mixture is a trick which can have only a FAPP (for all practical purposes) validity - to use a specification due to Bell \cite{Bell4}.
\item There is also a much more serious objection to the now described game. Most of the proponents of this solution to the puzzling aspects of linear superpositions of macroscopically distinguishable states seem to ignore a fundamental fact, i.e. that, within quantum mechanics, the correspondence between statistical ensembles and statistical operators is infinitely many to one. Thus, even if one would accept that the statistical operator which must be used is the one of Eq.(14), one has no reason to interpret it as describing the statistical ensemble  at the r.h.s. of Eq.(15). In fact, just to mention a trivial example, one immediately realizes that, e.g.,  also the following statistical mixture of three pure quantum states:
\end{itemize}

\begin{equation}
\mathcal E\{\frac{1}{\sqrt{2}}[\vert M\rangle+\vert N\rangle],\frac{1}{\sqrt{2}}[\vert M\rangle-\vert N\rangle],\vert M\rangle; \vert\beta\vert^{2}, \vert\beta\vert^{2}, (\vert\alpha\vert^{2}-\vert\beta\vert^{2})\},
\end{equation}
corresponds precisely to the same statistical operator of Eq.(14), but it still contains  embarassing superpositions of macroscopically distinguishable states.

Actually, the decoherence approach to the measurement problem has been repeatedly criticized (see, e.g. \cite{Adler} ) and, moreover,  the most serious proponents of this solution have plainly recognized that the  situation which we have outlined is puzzling, and have proposed  (vague) ways out from the puzzle. The most paradigmatic example  can be found in an important paper \cite{Joos} by Joos and Zeh,  which clearly admidts the arbitrariness of choosing one statistical mixture (the most reasonable from a perceptual point of view and the one corresponding to wave packet reduction), among the infinitely many which are possible: 

\begin{itemize}
\item {\it The local description is assumed and the specific choice of a basis can perhaps  be justified by a fundamental underivable assumption about the local nature of the observer ... and his way of perceiving,}
\item {\it No unitary treatment of the time dependence can explain why only one of these
dynamically independent components is experienced,}
\item {\it The difficulty in giving a complete derivation of classical concepts may as well signal
the need of entirely novel concepts.}
\end{itemize}

As a final remark we would like to stress that the most recent attempts to achieve  important technological improvements based on quantum mechanics - such  as Quantum Cryptography, Quantum Teleportation and Quantum Computation -  deal unavoidably with individual physical systems and they make a systematic use of wave packet reduction at the individual level as an important resource. Unless one ignores these fundamental aspects of modern research one has to face  the necessity of breaking the linear nature of quantum mechanics at an appropriate stage.

\section {The primitive ontology of a theory}
Many scientists maintain that the purely technical, formal and logical aspects of a theory represent all what deserves attention. We share with J S Bell and many others the opinion that further requirements must be imposed to any theoretical scheme  to be considered as a fundamental account of natural processes. We do not want to spend many words on this point; for a deep analysis  we refer the reader to a recent lucid paper \cite{Allori} in which the demand for  a  richer elaboration of the meaning of the formal scheme one is considering has been put forward. The authors of this paper have stressed the necessity of equipping any theory with what they call `the Primitive Ontology' (PO) of the formalism. In brief, the PO consists in the clear and precise specification of what the theory is fundamentally about.
	
	As already stated we will go quickly through this point, limiting ourselves to illustrate it with reference to two theories, two proposals for the solution of the macro-objectification problem which, at least at the nonrelativistic level, are fully consistent. We have  in mind Bohmian Mechanics and the Dynamical Reduction Models, in particular the so-called GRW theory. Their different status deserves to be stressed.  In fact while Bohmian Mechanics is, by assumption, predictively identical to quantum mechanics  the GRW theory represents a modification of the standard theory. They correspond to the two alternatives which Bell \cite{Bell3} has indicated as the only two possible ways out of the macro-objectification problem:
\begin {quote}
{\it Either the  wavefunction, as given by the Schr\"odinger equation, is not everything, or it is not right.}
\end{quote} 

Let us come to analyze these theories and to discuss their primitive ontologies.

\subsection{Bohmian Mechanics}
The formal aspects of the  proposal put forward in his celebrated papers \cite{Bohm} by D Bohm in 1952 can be summarized as follows:
\begin{itemize}
\item The quantum description of the state of a physical system is incomplete; for a complete specification one has to add to the wavefunction also the positions of all particles which constitute the system under examination (the positions are the hidden variables of the theory because we are able to prepare any desired wavefunction, but  we have no way to control at which point of the support of the wavefunction the particle actually is - and the theory assumes that it has a precise position),
\item The initial state of a system is fully specified by the assignment of the initial wavefunction in configuration space $\Psi(q_{1},q_{2},...,q_{n},0)\equiv\Psi(q_{i},0)$, and the initial positions $Q_{i}(0), (i=1,2,...,n)$  actually occupied by the constituents particles,
\item There are two linked evolution equations in the theory. The wavefunction evolves according to the standard quantum equation. Then one defines, in terms of the wave function,  a velocity field $v_{B,k}(Q_{j},t)$ for each particle  and considers a first order (in time) differential equation for the positions themselves. In formulae:

\begin{eqnarray}
i\hbar\frac{\partial\Psi(Q_{i},t)}{\partial t} & = & H \Psi(Q_{i},t);\;\;\frac{dQ_{k}(t)}{dt}=v_{B,k}(Q_{j},t);\\ \nonumber v_{B,k}(Q_{j},t) & = & \frac{\hbar}{m_{k}}\frac{Im[\Psi^{*}(Q_{j},t)\nabla_{k}\Psi(Q_{j},t)]}{\vert\Psi(Q_{j},t)\vert^{2}}.
\end {eqnarray}
\end{itemize}
The theory exhibits some nice features. First of all, in virtue of the continuity equation satisfied by Schr\"odinger's  wavefunction one immediately proves that if one defines the position density distribution $\rho(Q_{k},t)=\vert\Psi(Q_{k},t)\vert^{2}$ one has the so called equivariance property:
\begin{equation}
\rho(Q_{k},0)=\vert\Psi(Q_{k},0)\vert^{2}\Rightarrow \rho(Q_{k},t)=\vert\Psi(Q_{k},t)\vert^{2}.
\end{equation}
This equation tells us that if one has an ensemble of particles whose initial positions are distributed according to the modulus square of the initial wavefunction, then, by letting the particles  evolve along the uniquely defined trajectories ensuing from the differential equations for the positions, one recovers at any time the position density distributions predicted by quantum mechanics.

The theory is purposedly equivalent to standard quantum mechanics but it can be shown to imply that macroscopic objects end up, with the correct probabilities, in those positions that are assigned to them by the standard theory enriched with the reduction postulate. In short the theory solves the measurement problem in a perfectly consistent way.

The primitive ontology of the theory emerges clearly from the given picture:  all particles of the Universe have at all times perfectly defined positions, they move along precise trajectories in such a way to reproduce the position density distribution of standard quantum mechanics. In the case of micro-macro interactions, macroscopic systems end up, in turn, in the precise positions predicted by quantum mechanics plus the wave packet reduction postulate. Obviously, one has to assume that all measurements reduce essentially to position measurements, just as it happens when we consider a system with a macroscopic pointer, from the position of which we can infer the outcome of the measurement.   What the theory is about are the positions and they match perfectly  our perceptions concerning macroscopic systems. Using Bell's terminology,   the positions are {\it the  beables} of the theory, and everything is built up from them.

\subsection{The dynamical reduction program}
This attempt to account for all natural processes in terms of a unique dynamical principle is based on two important remarks: first of all the characteristic trait which makes the standard evolution and wave packet reduction radically different is that while the first is linear and deterministic, the second one is nonlinear and  stochastic. Secondly, if one wants to have a mechanism leading  to  the macro-objectification of some variable, one has to make a precise choice about it since different quantities are associated to noncommuting operators and one cannot ask that incompatible observables become simultaneously objective.  The most natural candidate, if one takes into account our definite perceptions, are the positions, in accordance with the remark by Einstein \cite{Einstein}:
\begin{quote}
{\it A macro-body must always have a quasi-sharply defined position in the objective description of reality.}
\end{quote}

Accordingly, in the first model of this kind which has been proposed \cite{ghirardi} (see also the comprehensive review\cite {bassi2}) one adds to Schr\"odinger's equation nonlinear and stochastic terms affecting all elementary constituents of any system and involving a universal localization mechanism. Formally one assumes that, at random times, with an appropriate frequency $\lambda\approx 10^{-16}sec^{-1}$, the wavefunction suffers a process corresponding to a localization of one of the constituents. When this process involves the $i-th$ particle the wavefunction changes according to:
\begin{eqnarray}
\Psi(q_{1},q_{2},...,q_{n})& \Rightarrow & \frac{\Lambda_{i}(x)\Psi(q_{1},q_{2},...,q_{n})}{\parallel\Lambda_{i}(x)\Psi(q_{1},q_{2},...,q_{n})\parallel}\\ \nonumber
\Lambda_{i}(x) & = & \frac {1}{(2\pi\sigma^{2})^{3/2}} \cdot e^{-\frac{(q_{i}-x)^{2}}{2\sigma^{2}}};\;\;\;\sigma\approx 10^{-5}cm.
\end{eqnarray}
A last specification concerns the positions at which collapses occur. It is embodied in the assumption that the position $x$ is chosen at random with a probability distribution:
\begin{equation}
dP(x\in [x, x+dx]\vert \Psi (t),i)=\parallel \Lambda_{i}(x)\Psi(q_{1},q_{2},...,q_{n})\parallel^{2}dx.
\end{equation}

The model exhibits some physically interesting features:
\begin{itemize}
\item The spatial probability density of the localizations, according to Eq.(20), practically coincides with the one which the standard theory attributes to position measurements,
\item It is immediately proved, by passing to the centre-of-mass and relative coordinates, that the theory entails the so called {\it Trigger Mechanism}: the localizations of the c.o.m. are amplified with the number of particles, actually of nucleons (because the frequency $\lambda$ is assumed to be proportional to the mass of the constituents), the value given above referring to nucleons,
\item According to this prescription a nucleon (or an atom) suffers a localization about every $10^{8}$ years, and, consequently, microscopic systems are  practically unaffected, while a macroscopic object, containing an Avogadro's number of particles, suffers a localization about every $10^{-7}$sec.,
\item The statistical operator obeys an equation of the Quantum Dynamical Semigroup type,
\item The theory is, in principle, testable against quantum mechanics.
\end{itemize}

The conclusion should be obvious: the proposed universal dynamics leaves all quantum predictions for microsystems unaltered but it accounts for wave packet reduction with probabilities in agreement with the quantum ones and for the classical behaviour of macroscopic systems, as well as for our definite perceptions with them.

In the simplified version we have presented the theory suffers of a serious limitation: the localizations break the symmetry character of the wavefunction for identical particles. This feature is easily corrected as firstly shown in\cite{Pearle}. Before coming to consider the ontology of the model we mention that much more formally elegant (even though physically equivalent) variants of the theory have been presented \cite{Pearle, ghirardi2}. They are based on the formalism of stochastic differential equations of the Ito or Stratonovich type.

Let us now come to discuss the PO of the theory. A first proposal has been put forward by Bell himself in his presentation of the GRW theory at the Imperial College meeting celebrating the centenary of Schr\"odinger's \cite{Bell3}:

\begin{quote}
{\it There is nothing in the theory but the wavefunction. ... However, the GRW jumps (which are part of the wavefunction, not something else) are well localized in ordinary space. Indeed each is centred on a particular spacetime point ({\bf x},t). So we can propose these events as the basis of the `local beables' of the theory. These are the mathematical counterparts in the theory to real events at definite times and places in the real world. ... A piece of matter  then is a galaxy of such events.}
\end{quote} 

This  assumption that `what the theory is fundamentally about' are the localizations themselves has been characterized as `The flashes ontology' in refs.\cite {Allori} and  \cite{Tumulka,Tumulka2,Tumulka3}. An alternative ontology has been proposed in \cite{Benatti} and has been subsequently characterized as `the mass density ontology'. This assumes that what the theory is about, what is real `out there', is the mass density in 3-dimensional space:
\begin{equation}
m({\bf x},t)=\sum_{i=1}^{N}\int_{R^{3N}} d{\bf r}_{1},...,d{\bf r}_{N}\delta({\bf r}_{i}-{\bf x})\vert\Psi({\bf r}_{1},...,{\bf r}_{N},t)\vert^{2}.
\end{equation}

This concludes our description of  two ways out from the  macro-objectification problem. As already mentioned there have been many other interesting attempts aiming at\ the same result. We have concentrated our attention on these two because they  are precise  models which, at the nonrelativistic level, account on the basis of a unique universal dynamics both for the behavior of microscopic systems, as well as for  the reduction process which takes place during measurement procedures. Bell has qualified them \cite{Bell4} as `exact', a term by which he intended {\it which are precisely formulated and which neither require nor are embarrassed by a conscious observer}. Bohmian Mechanics has been discussed many times and there are also entire books devoted to it. Concerning the GRW theory there is an extensive literature (which can be found, e.g., in \cite{bassi2}) in which many physically interesting aspects of the theory are analyzed.

\section{The recent debate}
In what follows we will be specifically interested in reviewing the new trends in the foundational investigations about quantum mechanics. In spite of many new proposals, we witness a revival of the old positions taken by the Copenhagen school in the thirties. In particular some scientists have expressed various criticisms concerning the two approaches we have just discussed and their consideration will allow us also to call attention to the new trends in the field.

\subsection{The case of Bohm theory}

A quite widespread critical attitude towards Bohmian Mechanics makes reference to the fact that since this theory is empirically indistinguishable from the standard theory, it should be considered an example of `bad science',  a `degenerate research program' in the sense of I Lakatos. To stress this position I cannot do better than quoting a sentence from a letter written in 1966 by S Weinberg to S Goldstein which can be found in ref.\cite{Hagar}:
\begin{quote}
{\it At the regular weekly luncheon meeting today ot our Theory Group, I asked my colleagues what they think of Bohm's version of quantum mechanics. The answers were pretty uniform and much what I would have said myself. First, as we understand it, Bohm's quantum mechanics uses the same formalism as ordinary quantum mechanics, including a wavefunction that satisfies the Schr\"odinger equation, but adds an extra element, the particle trajectory. The predictions of the theory are the same as for ordinary quantum mechanics, so, there seems  little point in the extra complication, except to satisfy some a priori ideas about what a physical theory should be like... In any case, the basic reason for not paying attention to the Bohm approach is not some sort of ideological rigidity, but much simpler -- it is just that we are all too busy with our own work to spend time on something that doesn' t seem likely to help us make progress with our real problems.}
\end{quote}

In our opinion it is remarkable that various (presumably) high level scientists and even a giant like S Weinberg do not consider important at least to mention the puzzling aspects of the standard interpretation of the theory, the aspects for which Bohm has been led to work out his proposal. Without paying the due attention to this fundamental point one cannot appreciate  the  great merit of the approach, i.e., the fact that it consistently solves the measurement problem and leads  to the classical behaviour of macroscopic objects, a feature which is by no means trivial.

At any rate, if the predictive equivalence of the two theories is considered as a drawback, people interested in foundational issues should not ignore that the other `exact' theory we have presented  qualifies itself as a rival theory of quantum mechanics which, in principle, can be tested against it. So, let us pass to analyze some of the critical remarks adressing the dynamical reduction program.

\subsection{The dynamical reduction models}
As already remarked, it is often  claimed that one can very well adopt the orthodox interpretation of quantum theory provided one takes into account the  decoherence due to the environment. Typically, J Bub, in describing what he calls `the new orthodoxy', claims \cite{Bub} that the superpositions of macroscopically different states actually occur but we do not see them due to the coupling with the environment. We have already discussed this aspect and we believe to have proved that, even though one might adopt it FAPP, one cannot avoid facing it when one is interested in the internal consistency of the theory. Moreover, invoking decoherence leads to consider practically only the statistical aspects of the theory: in a sense one claims that, when dealing with ensembles, one can use the reduced statistical operator obtained by disregarding the environmental degrees of freedom. But, as we have already remarked, modern technology deals more and more with individual physical systems and only if one resorts to individual reduction processes one can go on. We have the impression that the attitude of many of the adherents to the `decoherence' approach make an a priori and vague choice aimed to render nonrelativistic quantum mechanics  a theory which, from an experimental point of view, cannot be caught in telling a lie\footnote {For a brilliant discussion of the arguments presented in this subsection we refer the reader to ref.\cite {Hagar}.}.

Such, in our opinion, not well pondered attitude emerges also from a formal aspect to which these scientists make often reference: the so-called `ancilla argument'. It is well known that any quantum evolution equation of the quantum dynamical semigroup type for the statistical operator  is physically equivalent to a quantum mechanical theory with a unitary dynamics provided one considers an enlarged Hilbert space. The game is quite simple: one introduces a new quantum ancilla whose degrees of freedom are unaccessible. Then one cooks up a unitary dynamics in the Hilbert space $System\otimes Ancilla$ such that by partial tracing on the degrees of freedom of the ancilla one recovers the evolution equation for the statistical operator of a dynamical reduction model of the type we have considered. A natural question arises: what is the purpose of this position besides the desire to protect the standard formalism come what may? The ancilla has, by construction, no observable effect and the procedure amounts simply to introduce hidden elements whose only role is to save the formal structure of the standard theory. 

In view of the fact that dynamical reduction theories qualify themselves as rival theories of quantum mechanics, would it not be more serious, scientifically, to try, as R Penrose \cite{Penrose,Penrose2,Penrose3}, S Adler \cite{Adler2,Adler3,Adler4} and many others do, to investigate the possibility of performing some crucial test? I do not intend to suggest that the GRW theory has to be taken seriously in its present formulation as a fundamental theory of natural processes, but I cannot avoid calling attention to the fact that, being a precise theory, it can give some indications about where to look for possible violations of the linearity of quantum mechanics.

\section{Relativistic generalizations}
 Bell in one of his last papers, after having discussed Bohmian Mechanics and the GRW theory has stated \cite{Bell4}:
\begin{quote}
{\it The real problem now is which one of these two exact theories admits a relativistic generalization}
\end{quote}
Concerning this fundamental problem it is useful to mention that Bohmian Mechanics admits relativistic generalizations of various kinds, from the original attempts of Bohm and Hiley  \cite{Hiley} based on a preferred Lorentz frame, to the recent investigations by D\"urr {\it et al.} \cite{Duerr} and Salmos \cite{Salmos}, resorting to an arbitrary preferred space-time slicing. Berndl {\it et al.} \cite{Berndl} and  Dewdney and Horton \cite{Dewdney} suggest a preferred joint parametrization of the world lines. We are not interested here in the technical details of these approaches, what is relevant is that they are characterized by a general feature which turns out to be unavoidable: they are not  `genuinely invariant' in the precise sense that they require a (hidden) preferred reference frame. This is a consequence of the fact that \cite{Grassi} any theory which violates Bell's locality condition by violating the parameter independence requirement, does not admit a `genuine' relativistic generalization.

There have been also various attempts at a relativistic generalization of dynamical reduction models. The first is due to P Pearle \cite{Pearle2} and it has been discussed in detail and proved to be formally perfectly Lorentz invariant in ref.\cite{ghirardi5}. The idea is quite simple but deep: one considers a fermion field which is coupled to a meson field and introduces a reduction mechanism which forbids the superpositions of different mesonic states. Since the mesonic clouds associated to different positions of a fermion differ, one induces in this way an indirect localization of the fermions. However, a new problem arises: the introduction of stochastic processes in a relativistic context leads to the emergence of untractable divergences.

Other attempts should be mentioned, among them those of Dove and Squires \cite{Dove},  of  Dowker {\it et al.}  \cite{Dowker0,Dowker,Dowker1}, which are formulated on a discrete space-time. Up to very recent times no real step forward has been done. In the year 2006, R Tumulka has presented \cite {Tumulka,Tumulka2,Tumulka3} a relativistic generalization of the GRW theory for $N$ noninteracting distinguishable fermions based on the consideration of a multi-time Dirac equation. The theory sticks to what we have called the flashes ontology\footnote{Models based on the mass-density ontology require a serious reconsideration before one can resort to them for relativistic generalizations.}. It is particularly interesting to mention Tumulka's conclusions which he has reinforced recently \cite{Tumulka2}:
\begin{quote}
{\it A somewhat surprising feature of the present situation is that we seem to arrive at the following alternative: Bohmian mechanics shows that one can explain quantum mechanics, exactly and completely, if one is willing to pay with using a preferred slicing of space-time; our model suggests that one should be able to avoid a preferred slicing if one is willing to pay with a certain deviation from quantum mechanics.}
\end{quote}
\section{Quantum computation, nonlocality and realism}
Quite recently, an attitude which we see as an attempt to revive the Copenhagen position emerged in connection with two quite different subjects, i.e., on the one side, the investigations concerning the promising and interesting field of quantum computation and, on the other, some theoretical and experimental analyses of quantum nonlocality. Let us proceed to discuss them.

People working in quantum computation have repeatedly put forward the idea that what quantum mechanics is about is not `something existing out there', but only and exclusively `our information'. A paradigmatic example of this attitude appears in a paper by A Zeilinger \cite{Zeilinger}:
\begin{itemize}
\item {\it The distinction between reality and our knowledge of reality, between reality and information, cannot be made,}
\item Bell's result suggests\ {\it that the concept of reality itself is at stake,}
\item {\it There is no way to refer to reality without using the information we have about it.}
\end{itemize}

Leaving apart the tautological structure of the last sentence, I believe that the best response to the first has been given in a recent paper \cite{Daumer}:
\begin{quote}
{\it The distinction between reality and our knowledge of reality, not only can be made; it must be made if the notions of knowledge and information are to have any meaning in the first place.}
\end{quote}

To further illustrate the discussion about the recent positions taken by some scientists we consider it appropriate to mention a recent exchange of views between D Mermin and the present author \cite{Mermin,ghirardi4}. Mermin asserts that the present status of quantum computation is completely satisfactory on the basis of the fact that:
\begin {itemize}
\item {\it ... all gates, including the measurement gate, alter the state associated with the incoming Q-bits in a well defined, generally discontinuous manner which is precisely defined by the state,}
\item  {\it ... in spite of the fact that there remains the distinction between linear, invertible, unitary gates with output state fully determined by the imput, and nonlinear, irreversible, measurement gates, with output state stochastically determined by the input, ..., the action of both types of gates are fully determined, with nothing left to the discretion of the theoretical physicist.}
\end{itemize}

We perfectly agree on these statements, but we stress that they amount simply to a way of evading the real problem. In fact Mermin himself stresses that:
\begin{quote}
{\it it is through the readings of 1-Qbit measurement gates, and only through such readings, that one can extrect information from a quantum computer ... measurement is essential at both ends of a quantum computation.  ... the `user' which looks at a visual display or a print out} [plays the fundamental role in the process].
\end{quote}

Here one can see, on the one side, the plain recognition that the very practice of quantum information and computation requires to accept the validity of two different and incompatible dynamical principles. On the other side, one ignores completely  the fact that while we know everything about the working of the unitary gates used to  implement our algoritms, we have an extremely vague idea of why and when individual  physical processes occur which are not governed by the (usually) assumed universal laws of quantum mechanics and actually contradict it. This is, in essence the criticism to Mermin that we have raised. He has replied \cite {Mermin2}  and, besides making various  remarks he has made quite clear his position:
\begin{quote}
{\it I stress that the state and the unitary transformations it is subject to are mathematical abstractions that enable one to compute ... the probabilities of the readings of the final measurement gates. Mathematical abstractions do not require stochastic `hits' originating in unknown physical processes or interactions with gravitons to be reset; they are reset by us, when we acquire more information and want to calculate what we expect to experience next.}
\end{quote}

This sentence makes explicit reference to the GRW theory and to the proposals by Penrose of relating wave packet reduction to gravity. We believe that one cannot ignore that these alternative theories represent serious attempts to account consistently, in physical terms and not by vague verbal assumptions,  for what {\it `we will experience ... next'}.

To conclude this first part of our comments we would like to mention that the idea that quantum mechanics deals simply with `information', is not new. For instance it was put forward during the discussions concerning the decoherent histories approach to quantum mechanics when it has been suggested to look at them exclusivively with reference to  IGUSes ({\it information gathering and using systems}). Bell has considered this position and he has made clear \cite{Bell4} that he was inclined to reject any reference to information unless one would, first of all, answer to the following basic questions: {\it Whose information?, Information about what?}

This remark allows to focus in a precise way the two incompatible positions which characterize the present debate  which, in our opinion, echoes the one of more than 60 years ago. In fact, e.g., Mermin himself, has recently taken a quite precise position \cite {Mermin3} concerning the issue under discussion:
\begin{quote}
{\it The question of `information about what' is a fundamentally metaphysical question that ought not to distract tough-minded physicists.}
\end{quote}

\noindent This concludes our first set of remarks. 

We come now to the second point, i.e., to the recent debate concerning nonlocality. An anticipation  of this attitude is already contained \cite{Zeilinger} in the second sentence reported at the beginning of this section.  The story goes as follows: A Leggett has considered \cite{Leggett} a class of theories, referred to as {\it nonlocal realistic theories}, which exhibit nonlocal features but satisfy a particularly natural (in his opinion) set of conditions. Leggett himself has shown that, in spite of the fact that one could not derive a Bell's type of inequality for such theories, they were at variance with the predictions of quantum mechanics and had to be abandoned. Subsequently various authors \cite{Zeilinger2,Gisin,Zeilinger3,Colbeck} have considered the same problem both with reference to Leggett's proposal as well as for a somewhat extended class of nonlocal theories and have proved once more their incompatibility with quantum mechanics. Even experimental tests of Leggett's model have been performed, confirming the crash with quantum predictions.

This line of research is, in our opinion, quite interesting and deserves to be pursued. However, the above facts - i.e. the supposed `naturalness' of the requests and the persisting incompatibility with quantum mechanics - have given rise to a very peculiar type of argument. In a sketchy way it can be summarized as follows: the derivation of Bell's inequality is based on two assumptions: Locality {\it and} Realism. Accordingly, the experimental violation of Bell's inequality requires to abandon at least one of these assumptions. Since giving up locality along the `natural' way suggested by Leggett does not eliminate the crash of such proposals with quantum theory, the experimentally tested incompatibility   gives a clear indication that one has to give up the assumption of realism. This argument, combined with the already emerging opinion, coming from the quantum computational community, i.e. that {\it  the concept of reality itself is at stake}, has seen the adoption of this point of view by many, even extremely brilliant physicists \cite{Zeilinger,Zeilinger2,Aspect}.

We would like to make some remarks on this point. First of all we will call attention on  the factual irrelevance of the argument. Secondly we will stress, and we consider this an extremely important point, that the just mentioned papers take  a completely mistaken attitude towards the real meaning of Bell's work. Let us proceed to present the just mentioned criticisms.

\begin{itemize}
\item The class of models which have been considered is quite restricted, in particular it embodies a sort of weak locality request, which is in no way compelling. This point has been already raised in an important paper \cite{Colbeck}.
\item There is no doubt that there exist theories which can legitimately  be qualified as `realistic' and fully agree with the standard theory. We  mention first of all Bohmian Mechanics and, secondly, a quite elementary toy example dealing with a system of two entangled spin $1/2$ particles in the singlet state which has been devised in 1964 by Bell in his fundamental paper \cite{Bell1}.
\end{itemize} 
   
In agreement with these remarks one cannot claim that the program of building up a `fully realistic nonlocal theory' equivalent to quantum mechanics is not viable, an important counterexample of what has been suggested in the above mentioned papers. But this remark is by no way the most relevant one, actually it turns out to be rather marginal. What really matters is the fact that the derivation of Bell's inequality in no way whatsoever needs an assumption of realism. In spite of this  fundamental fact, which everybody can verify by going carefully through the proof, a large part of the scientific community shares the completely wrong opinion that realism is among the basic assumption needed for the derivation of Bell's result.   Bell himself has stressed this  aspect and has remarked  \cite{Bell5} that it is extremely difficult to eradicate this prejudice\footnote{It has to be remarked that deterministic hidden variable theories assume that the complete specification of the state of the system implies that all physical properties are actually possessed by the systems prior to any measurement process. This is equivalent to the request of realism discussed by  the above mentioned authors}: 
\begin{quote}
{\it My own first paper (Physics {\bf 1}, 195 (1965.) on this subject starts with a summary of the EPR argument {\bf from locality to} deterministic hidden variables. But the commentators have almost universally reported that it  begins with deterministic hidden variables.}
\end {quote}

This being the situation we must conclude that in no way whatsoever Bell's inequality has something to do with realism. It simply identifies in a straightforward and lucid way that what quantum phenomena impose to us is to accept the unescapable fact that natural processes involving entangled states of composite and far-away systems turn out to be unavoidably non-local. If one is keen to consider the possibility of giving up realism he can very well try his luck, but for sure such a big price is not necessary, and even not suggested, by the quantum nonlocal correlations. 

The question whether Bell's theorem involves some request of realism has been discussed and proven once more to be unjustified by various authors \cite{Norsen, Tausk, Laudisa}.

We would like to conclude this  analysis of some of the most relevant issues of the foundational investigations on quantum mechanics by expressing our worries concerning the fact that there are some indications that 80 years of lively debate on the conceptual problems of our best theory by the most brilliant physicists of the last century are facing a serious risk of being cancelled by those scientists who Mermin has qualified as the Q-computation crowd or by those who derive from experimental results inspired by not strictly convincing theoretical models unjustified conclusions concerning such an important issue as the one of the reality of the world around us.

\end{document}